# About empty waves, their effect, and the quantum theory

*Sofia Wechsler* [1)]

**Abstract**
When a quantum object – a particle as we call it in a non-rigorous way – is described by a multi-branched wave-function, with the corresponding wave-packets occupying separated regions of the time-space, a frequently asked question is whether the quantum object is actually contained in only one of these wave-packets. If the answer is positive, then the other wave-packets are called in literature *empty waves*. The wave-packet containing the object is called a *full wave*, and is the only one that would produce a recording in a detector.

A question immediately arising is whether the empty waves may also have an observable effect. Different works were dedicated to the elucidation of this question. None of them proved that the hypothesis of full/empty waves is correct – it may be that the Nature is indeed non-deterministic and the quantum object is not confined to one region of the space-time. All the works that proved that the empty waves have an effect, in fact, proved that *if* there exist full and empty waves, *then* the latter may have an observable effect.

This is also the purpose and the limitation of the present work. What is shown here is that if the hypothesis is true, the empty waves have an influence. An experiment is indicated which reveals this influence. The analysis of the experiment is according to the quantum formalism. This experiment has the advantage of being more intuitive and practically more feasible than a previous proposal also in agreement with the quantum formalism. However, the presently proposed experiment also shows that the quantum theory *is not in favor* of the above hypothesis.

**Abbreviations:**
DC = down-conversion.
rate = average number of events of a certain type per unit-time
UV = ultraviolet
w.f. = wave-function

## 1. Introduction

The wave-function (w.f.) of a quantum system was proved to be, by an endless series of experiments, an absolutely trustable tool for computing probabilities of experimental results. However, the probabilistic appearance of the properties of microscopic objects is regarded by many people with suspicion. The de Broglie-Bohm theory considers that the w.f. is some sort of a field which limits the movement of a particle, establishing in which regions of the space-time it is allowed to be, and in which not. Then, in each trial of the experiment the particle occupies one of the allowed regions. Therefore, this theory suggests that the microscopic world is though deterministic. A generalization of this view is the hypothesis of full/empty waves. This hypothesis says that if the w.f. consists in a superposition of wave-packets, one of them possesses a particular property which impresses a detector. This wave-packet is called a *full wave* and the other wave-packets, *empty waves*. The full and empty waves interfere, and everything goes according to the quantum formalism. Nothing can tell which wave-packets are full and which are empty, unless a detectors are placed on these wave-packets.

Then, a question immediately arises: may though the empty waves manifest somehow their presence? Different researchers considered that a positive answer to this question implies a conflict with the predictions of the standard quantum formalism, and proposed experiments to reveal such contradictions **[1, 2]** (see also **[3]** for a review of a couple of experiments and a critical discussion thereof). But the experiment showed that the quantum formalism is correct, see for instance **[4]**. In fact, no experiment is known until today, which contradicts the quantum formalism.

―――――――――――――――――
[1)] Computer Engineering Center, P.O.B 2004, Nahariya, 22265, Israel.

To the difference from the above works, L. Hardy described a thought-experiment **[5]** whose analysis points to an effect of the empty waves, while being in full agreement with the quantum formalism. That, on one condition, *that the empty waves exist*. Hardy didn't prove the correctness of the hypothesis of full/empty waves, what he proved was that *if* this hypothesis is correct, *then*, the empty waves have an observable effect. To be clear, his experiment may be interpreted in an alternative way, that refutes realism. Measurement results may be considered a consequence of the measurements only, i.e. there may be no difference between the wave-packets before the measurement.

The present text proposes another experiment that shows an effect of the empty waves, again, in the assumption that they exist. As in **[5]**, the analysis and predictions of this experiment comprise no disagreement with the quantum formalism. However, this experiment is simpler than the proposal in **[5]**, and practically more feasible. Actually a variant of it was already performed, **[6]**, but with a different configuration.

The present proposal relies on the behavior in time of a 2particle interference effect. Pairs of photons landing on a non-linear crystal interfere with identical pairs born in the crystal by down-conversion (DC). When the interference is constructive, the flux of pairs exiting the crystal is increased. However, the number of pairs coming to the crystal is too small, and the presence of a pair in the crystal too brief, for creating the required interference. Thus, most of pairs generated in the crystal seem to be born when no incident pairs are present. Then the question arises between whom and whom occurred the constructive interference? The explanation may be that while the flux of full waves – the detectable pairs – is discrete in time and each pair of short duration, a continuous flux of empty waves incident to the crystal and a continuous flux of empty waves born in the crystal, allow the above interference.

The rest of the text is organized as follows. Section **2** describes the proposed experiment. Section **3** discusses the enhanced emission of pairs in the light of the full/empty waves assumption. Section **4** comprises a critical discussion of this assumption. Additional mathematical details are given in Appendix.

## 2. A down-conversion experiment

Two identical non-linear crystals $X'$ and $X$ are illuminated by identical coherent beams of UV-photons, emitted, respectively, one by a source $L'$ and by a source $L$, fig. 1. We follow here the evolution of one UV-photon. In each crystal takes place down-conversion to signal-idler pairs

$$(1) \quad |1_{p'},0,0\rangle \rightarrow \beta|1_{c'},0,0\rangle + \sum_{j,k} \alpha(j',k')|0,1_{j'},1_{k'}\rangle, \quad |\beta|^2 + \sum_{j,k} |\alpha(j',k')|^2 = 1$$

$$(2) \quad |1_p,0,0\rangle \rightarrow \beta|1_c,0,0\rangle + \sum_{j,k} \alpha(j,k)|0,1_j,1_k\rangle, \quad |\beta|^2 + \sum_{j,k} |\alpha(j,k)|^2 = 1.$$

The notation $|\ell_q, m_r, n_s\rangle$ describes the state of $\ell$ UV-photons with wave-vector $q$, of $m$ signal-photons with wave-vector $r$, and $n$ idler-photons with wave-vector $s$. The summations are over the signal and idler wave-vectors that satisfy the phase matching conditions, see the Appendix, Part 1.

Behind the crystals, two identical screens, $E'$ and $E$, let pass through two tiny holes only DC-pairs with definite directions of flight and wavelengths – let's denote the wave-vectors in such a pair by $j_0$ for the signal and $k_0$ for the idler. We will refer to such pairs as "selected". Therefore eqs. (1) and (2) become

$$(3) \quad |1_{p'},0,0\rangle \rightarrow \beta|1_{c'},0,0\rangle + \alpha(j_0,k_0)|0,1_{j_0},1_{k_0}\rangle + \sum_{\substack{j' \neq j_0 \\ k' \neq k_0}} \alpha(j',k')|0,1_{j'},1_{k'}\rangle,$$

$$(4) \quad |1_p,0,0\rangle \rightarrow \beta|1_c,0,0\rangle + \alpha(j_0,k_0)|0,1_{j_0},1_{k_0}\rangle + \sum_{\substack{j \neq j_0 \\ k \neq k_0}} \alpha(j,k)|0,1_j,1_k\rangle,$$

$$(5) \quad |\beta|^2 + |\alpha(j_0,k_0)|^2 + \sum_{\substack{j' \neq j_0 \\ k' \neq k_0}} |\alpha(j',k')|^2 = |\beta|^2 + |\alpha(j_0,k_0)|^2 + \sum_{\substack{j \neq j_0 \\ k \neq k_0}} |\alpha(j,k)|^2 = 1.$$

**Experiment A)** Behind the screens $E$' and $E$ are placed pairs of detectors, $S$', $I$', respectively $S$, $I$. If $N_0$ is the rate of *single* UV-photons from each source, (average number per unit-time), then according to (3) and (4) the rate of detections in coincidence in the two pairs of detectors and in total, will be

(6) $<Q'> = <Q> = N_0|\alpha(j_0, k_0)|^2$,   $<Q_T> = 2<Q'> = 2<Q> = 2N_0|\alpha(j_0, k_0)|^2$.

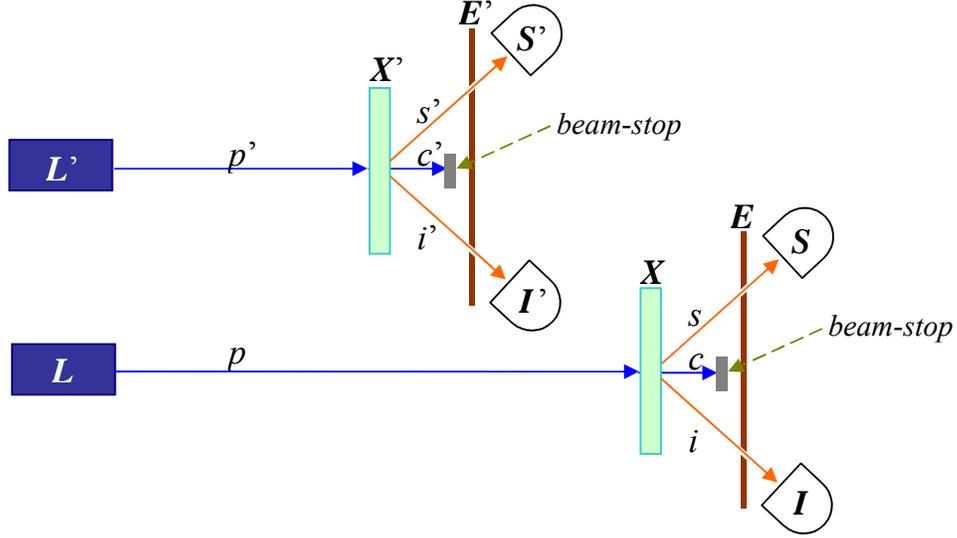

**Figure 1.** DC-pair production in two non-linear crystals.

**Experiment B)** The detectors $S$', $I$' are removed, fig. 2, and the selected pairs from $X$' are deflected by means of mirrors to the crystal $X$. In the crystal $X$' the process doesn't change, but in the crystal $X$ interference occurs between the amplitude of probability of the pair coming from $X$' and the amplitude of probability of the pair born in $X$. For this phenomenon to be observable, in what follows we consider only the detections during intervals of time when there is constant relative phase between the beams emitted by the two sources. To shorten the formulas, let's denote by φ the difference between the phase of the photon $p$ on the input face of the crystal $X$ and the phase of the photon $p$' on the input face of the crystal $X$'. The system state on the input face of $X$ is then

(7) $|\Phi> = e^{\iota\sigma}\alpha(j_0, k_0)|0, 1_{j_0}, 1_{k_0}> + e^{\iota\varphi}|1_p, 0, 0>$.

This expression is normalized so as to indicate the same probability for a selected pair from $X$', as in (3), and probability 1 for having a UV-photon on the path $p$, therefore *the norm of $|\Phi>$ is more than* 1.
σ is the phase acquired by this pair during the flight from $X$' to $X$, $\sigma = \ell_s |j_0| + \ell_i |k_0|$, where $\ell_s$ and $\ell_i$ are the path-lengths of the signal respectively idler photon.
Inside the crystal $X$ the selected pair coming from $X$' undergoes the transformation

(8) $\hat{U}|0, 1_{j_0}, 1_{k_0}> = -\alpha^*(j_0, k_0)|1_c, 0, 0> + [1 - \gamma(j_0, k_0)]|0, 1_{j_0}, 1_{k_0}> - \sum_{\substack{j \neq j_0 \\ k \neq k_0}} \gamma(j, k)|0, 1_j, 1_k>$,

(see the derivation of this expression in Part 1 of the Appendix, expression (A10) ).
Introducing in (7) the transformation (8) and the transformation (4) undergone in $X$ by the photon $p$, one gets

(10) $|\Phi\rangle = [-e^{\iota\sigma}|\alpha(j_0, k_0)|^2 + e^{\iota\varphi}\beta]|1_c, 0, 0\rangle + \alpha(j_0, k_0)\{e^{\iota\sigma}[1 - \gamma(j_0, k_0)] + e^{\iota\varphi}\}|0, 1_{j_0}, 1_{k_0}\rangle$

$$+ \sum_{\substack{j \neq j_0 \\ k \neq k_0}} [e^{\iota\varphi}\alpha(j, k) - e^{\iota\sigma}\alpha(j_0, k_0)\gamma(j, k)]|0, 1_j, 1_k\rangle.$$

As explained in the Appendix, see (A11), the value $\gamma(j_0, k_0)$ is extremely small in comparison with 1. Note in fig. 2 that the crystal $X$ is rotated so as to make the paths of the photons coming from $X'$, overlap with the paths of the corresponding photons born in $X$. In this way, in the crystal $X$ and beyond it, nothing reminds the origin of a selected pair. That facilitates 2photon interference.

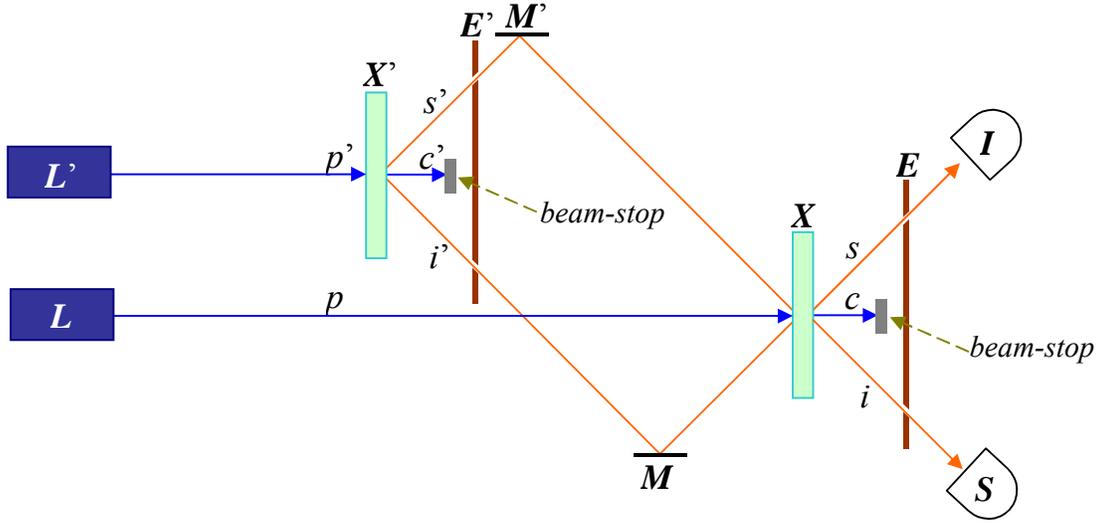

**Figure 2.** Bringing the paths of the DC-photons from two crystals to overlap.

Two cases are particularly interesting:

a) $\sigma = \varphi + \pi$. The expression (10) shows that since $\gamma(j_0, k_0)$ is negligibly small the selected type is quenched, practically no selected pairs can be found beyond $X$. This crystal stops generating such pairs. A more detailed calculus in Part 2a of the Appendix, proves that the pairs coming from $X'$ are up-converted in $X$ to UV-photons.

b) $\sigma = \varphi$. The selected pair amplitude in (10) becomes $2e^{\iota\varphi}\alpha(j_0, k_0)$, yielding for these pairs the rate,

(11) $\langle Q_E \rangle = 4N_0|\alpha(j_0, k_0)|^2$,

where the subscript $E$ means enhanced. Indeed, comparing (11) with (6) one can see that $\langle Q_E \rangle = 2\langle Q_T \rangle$, i.e. $\langle Q_E \rangle$ is twice the rate of selected pairs obtained in the experiment **A** from $X'$ and $X$ together.

Two components contribute to $\langle Q_E \rangle$. A part $\langle Q' \rangle = \frac{1}{4}\langle Q_E \rangle$ comes from $X'$, see experiment **A**. The rest of $\frac{3}{4}\langle Q_E \rangle$ is produced in $X$ as proves Part 2b in the Appendix. Out of this, $\frac{1}{4}\langle Q_E \rangle$ would be produced in $X$ anyway if there were no contribution from outside, as showed the experiment **A**. Therefore the extra-rate of $\frac{1}{2}\langle Q_E \rangle$ is a pure result of the constructive interference.

The implications of the case b are further analyzed in the next section.

## 3. The enhanced emission and empty waves

The problem posed in this text is, between whom and whom takes place this interference? We are going to show below that the number of selected pairs coming from *X'* is not enough for producing the interference necessary for the enhanced emission.

The very low value of $|\alpha|^2$ which is $\sim 10^{-11}$ entails the following limitation

(12) $T = \langle Q_E \rangle^{-1} \gg \delta$.

*T* is the mean interval of time between two consecutive selected pairs exiting the crystal *X*. The mean interval of time between the arrival at *X* of two consecutive pairs from *X'* is 4*T*. $\delta$ is the maximal interval of time during which a DC-pair may be present in *X*. If the inequality (12) is correct, then, a pair from *X'* has practically no chance to be present in *X* when a selected pair is born in the latter, see fig. 3. This is why the question arises, between whom and whom takes place the interference.

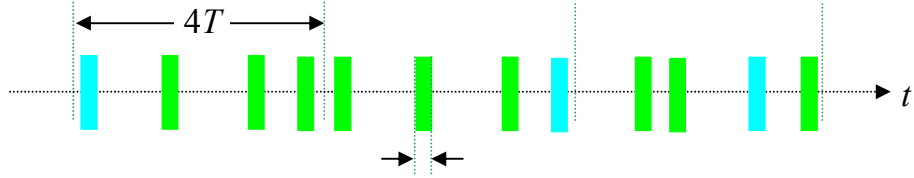

**Figure 3.** The presence of selected pairs in the crystal *X* during the time. Bluish rectangles symbolize pairs coming from *X'*, green rectangles symbolize pairs born in *X*. The colors are only for guiding the eye.

The order of magnitude of *T* may be for example $10^{-3}$ seconds. $\delta$ is the biggest between the following two values:

1) $\tau_{pcoh}$, the pair-coherence-time or in other words the time-duration of a DC-pair; it is of the same order of magnitude as the coherence time of each photon in the pair, $\sim 10^{-13}$ seconds, see for details ref. **[7]**.
2) $nD_{max}/c$, where *n* is the refraction index of the crystal, $D_{max}$ is the maximal path-length of a DC-photon in the crystal, and c the light velocity. For a $LiIO_3$ crystal as long as 1.5cm, $nD_{max}/c \sim 10^{-10}$ seconds.[2)]

Then $\delta \sim 10^{-10}$ seconds, and the inequality (12) is correct. A selected pair coming from *X'* leaves the crystal *X* much before a new pair is born in *X*.

The full/empty waves hypothesis offers the following solution: it is known that the train-wave of the DC-pairs has a duration as long as the coherence-time of the UV-beam, which may be tuned to be longer than *T*. This train-wave is continuous in time, not discrete. What is discrete is the occurrence of the full waves, those pairs that are detected if detectors are introduced. The rest of the train is empty waves.
Thus, the input face of the crystal *X* is continuously illuminated by the train-wave of pairs from *X'*. Also from *X* there emerges a continuous train-wave of pairs, out of which a discrete part is full as showed the calculi above. The interference is therefore between the two continuous train-waves. The fact that some pairs are full waves while most of the train-wave is empty waves, has no relevance in interference. As said in section **1**, the full/empty waves hypothesis assumes that the full and the empty waves interfere and everything goes according to the quantum formalism.

---
[2)] The value of *n* is different for the signal and the idler, and so is $D_{max}$ because the two photons fly at different angles with respect to the crystal surface; however, for the comparison in (11) is sufficient the order of magnitude of $nD_{max}/c$.

## 4. Discussion

As said in section **1**, the above argument that empty waves have an effect is valid only if the full/empty waves assumption is itself valid. In the present study, the DC transformation, eq. (A6), shows that a UV-photon state transforms into a quantum superposition of a UV-photon state and a signal-idler state (the latter having a very small amplitude). A quantum superposition is *not a mixture*, and there is no hint that in a given trial of the experiment one term of the superposition is a full wave and the other term an empty wave. Moreover, the quantum theory seems to disfavor this assumption.

Consider the experiment **B**. From the calculi in Part 2 of the Appendix there results that the difference between the UV-photon rates on the paths $p$ and $c$ is exactly matched by the difference between the rate of the pairs exiting the crystal $X$ and the rate of the pairs coming from $X$', see eqs. (A13), (A15), (A16). The equality between the rate of UV-photons lost on down-conversion and the rate of generated pairs, is a confirmation of the energy conservation.

Experimentally it is very difficult to confirm this equality since the down-conversion rate is so low that it is hard (maybe impossible) to measure the difference between the photon rates on $p$ and on $c$. Then let's rely on the fact that the theory predicts this equality. Another, strange situation appears, as follows.

These rates are averages, s.t. the above equality confirms the energy conservation only on average. But the energy conservation should be fulfilled in each individual trial of the experiment. In the terminology of full/empty waves, each full wave containing a DC-pair and exiting $X$, should either have come as a full wave from $X$', or have resulted from the down-conversion of a full wave in the UV-photon train-wave passing through $X$. Here the quantum theory imposes a prohibition: it is not possible to put detectors on the paths $s$', $i$' because that would decohere the w.f. (7) destroying the constructive interference that leads to the result (11). The paths $s$', $i$', and $p$ cannot be tested in the same trial of the experiment as the paths $s$, $i$, and $c$. Thus, in this experiment one cannot test if the energy conservation is fulfilled per individual trial.

## Appendix

## Part 1

The down-conversion and up-conversion Hamiltonian is

(A1) $\hat{H} = \sum \{g(\boldsymbol{k}_{UV}, \boldsymbol{j}, \boldsymbol{k})\, \hat{a}_{UV}\, \hat{a}^{\dagger}_j\, \hat{a}^{\dagger}_k + g^*(\boldsymbol{k}_{UV}, \boldsymbol{j}, \boldsymbol{k})\, \hat{a}^{\dagger}_{UV}\, \hat{a}_j\, \hat{a}_k\}.$

The summation is over all the wave-vectors $\boldsymbol{j}, \boldsymbol{k}$, that satisfy the phase matching conditions; $g(\boldsymbol{k}_{UV}, \boldsymbol{j}, \boldsymbol{k})$ is a coupling constant dependent on the nonlinear susceptibility; $\boldsymbol{k}_{UV}$ is considered here fixed and it will be omitted below in the argument list of the functions.

The unitary transformation that a system undergoes under the interaction described by the Hamiltonian $\hat{H}$, is $\hat{U} = \exp(\iota \hat{H} t/\hbar)$. Let's denote $\eta(\boldsymbol{j}, \boldsymbol{k}) = g(\boldsymbol{j}, \boldsymbol{k}) t/\hbar$. This quantity is dimensionless.

Let's develop the operator $\hat{U}$ in powers of the exponent

(A2) $\hat{U} = 1 + \iota \sum\limits_{j,\,k} (\eta\, \hat{a}_{UV}\, \hat{a}^{\dagger}_j\, \hat{a}^{\dagger}_k + \eta^*\, \hat{a}^{\dagger}_{UV}\, \hat{a}_j\, \hat{a}_k)$

$\quad - \tfrac{1}{2} \sum\limits_{j,\,k} (\eta\, \hat{a}_{UV}\, \hat{a}^{\dagger}_j\, \hat{a}^{\dagger}_k + \eta^*\, \hat{a}^{\dagger}_{UV}\, \hat{a}_j\, \hat{a}_k) \sum\limits_{j',\,k'} (\eta\, \hat{a}_{UV}\, \hat{a}^{\dagger}_{j'}\, \hat{a}^{\dagger}_{k'} + \eta^*\, \hat{a}^{\dagger}_{UV}\, \hat{a}_{j'}\, \hat{a}_{k'})$

$\quad - (\iota/3!) \sum\limits_{j,\,k} (\eta\, \hat{a}_{UV}\, \hat{a}^{\dagger}_k\, \hat{a}^{\dagger}_k + \eta^*\, \hat{a}^{\dagger}_{UV}\, \hat{a}_k\, \hat{a}_k) \sum\limits_{j',\,k'} (\eta\, \hat{a}_{UV}\, \hat{a}^{\dagger}_{j'}\, \hat{a}^{\dagger}_{k'} + \eta^*\, \hat{a}^{\dagger}_{UV}\, \hat{a}_{j'}\, \hat{a}_{k'})$

$\quad \times \sum\limits_{j'',\,k''} (\eta\, \hat{a}_{UV}\, \hat{a}^{\dagger}_{j''}\, \hat{a}^{\dagger}_{k''} + \eta^*\, \hat{a}^{\dagger}_{UV}\, \hat{a}_{j''}\, \hat{a}_{k''})$

$\quad + \ldots .$

Applying the transformation $\hat{U}$ to one UV-photon, and recalling that $\hat{a}_j \hat{a}_k |1,0,0\rangle = \hat{a}_{UV} |0,1_j,1_k\rangle = 0$, and also $\hat{a}_j \hat{a}_k |0,1_{j'},1_{k'}\rangle = 0$, one gets

(A3) $\hat{U}|1,0,0\rangle = [1 - \frac{1}{2}\xi^2 + (4!)^{-1}\xi^4 - (6!)^{-1}\xi^6 \pm ...]|1,0,0\rangle$

$$+ \iota \sum_{j,k} \eta(j,k)[1 - (3!)^{-1}\xi^2 + (5!)^{-1}\xi^4 - (7!)^{-1}\xi^6 \pm ...]|0,1_j,1_k\rangle$$

$$= \cos\xi |1,0,0\rangle + (\iota/\xi) \sin\xi \sum_{j,k} \eta(j,k),$$

where

(A4) $\xi = \sqrt{\sum_{j,k} |\eta(j,k)|^2}$.

The following notations will be used below

(A5) $\alpha(j,k) = \iota \eta(j,k)[1 - (3!)^{-1}\xi^2 + (5!)^{-1}\xi^4 - (7!)^{-1}\xi^6 \pm ...] = (\iota/\xi) \eta(j,k) \sin\xi,$ (a)

$\beta = 1 - \frac{1}{2}\xi^2 + (4!)^{-1}\xi^4 - (6!)^{-1}\xi^6 \pm ... = \cos\xi,$ (b)

$\gamma(j,k) = \eta^*(j_0, k_0) \eta(j,k)[\frac{1}{2} - (4!)^{-1}\xi^2 + (6!)^{-1}\xi^4 \pm ...],$ (c)

Using (A5a) and (A5b) the transformation (A3) becomes

(A6) $\hat{U}|1,0,0\rangle = \beta|1,0,0\rangle + \sum_{j,k} \alpha(j,k)|0,1_j,1_k\rangle$.

One can see from (A5a) and (A4) that

(A7) $\sum_{j,k} |\alpha(j,k)|^2 = \sin^2\xi,$

therefore

(A8) $\beta^2 + \sum_{j,k} |\alpha(j,k)|^2 = 1$.

The quantity $\sum_{j,k}|\alpha(j,k)|^2$ is extremely small, and in consequence $\beta$ is positive and very close to 1.

Applying the transformation $\hat{U}$ to one DC-pair, for instance to a selected pair, one gets

(A9) $\hat{U}|0,1_{j_0},1_{k_0}\rangle = (\iota/\xi) \eta^*(j_0, k_0) \sin\xi |1,0,0\rangle$

$$+ [1 - \frac{1}{2}|\eta(j_0, k_0)|^2 + (4!)^{-1}|\eta(j_0, k_0)|^2\xi^2 - (6!)^{-1}|\eta(j_0, k_0)|^2\xi^4 \pm ...] |0,1_{j_0},1_{k_0}\rangle$$

$$- \eta^*(j_0, k_0) \sum_{\substack{j \neq j_0 \\ k \neq k_0}} \eta(j,k)[\frac{1}{2} - (4!)^{-1}\xi^2 + (6!)^{-1}\xi^4 \pm ...] |0,1_j,1_k\rangle.$$

Using the notation (A5c) the transformation (A9) becomes

(A10) $\hat{U}|0, 1_{j_0},1_{k_0}\rangle = -\alpha^*(j_0, k_0)|1,0,0\rangle + [1 - \gamma(j_0, k_0)]|0,1_{j_0},1_{k_0}\rangle - \sum_{\substack{j \neq j_0 \\ k \neq k_0}} \gamma(j,k)|0,1_j,1_k\rangle.$

Note that while most of the pairs pass through the crystal, a fraction $|\alpha(j_0, k_0)|^2$ is up-converted, while another fraction $\sum_{j \neq j_0, k \neq k_0} |\gamma(j, k)|^2$ is re-converted to other combinations of signal and idler wavelengths.

Some additional useful relations result from (A5),

(A11) $\gamma(j, k) = \eta^*(j_0, k_0)\, \eta(j, k)\, (1 - \beta)/\xi^2 = \alpha^*(j_0, k_0)\, \alpha(j, k)\, (1 - \beta)/\sin^2\xi$

One can see that $\gamma(j, k)$ is of the order of magnitude of $|\alpha|^2$; since $|\alpha|^2$ is extremely small, $\gamma$ is extremely small too.

## Part 2

The two particular cases in the experiment **B** are analyzed here.

a) For $\sigma = \varphi - \pi$ the term with $|0, 1_{j_0}, 1_{k_0}\rangle$ in (10) practically vanishes. Besides the fact that no more selected pairs are generated in $X$, the pairs coming from $X'$ are up-converted to UV-photons.

To show this we will calculate the rate of photons on the path $c$ and compare it with the rate on the path $p$.

Introducing in (10) the relation $\sigma = \varphi - \pi$, and since according to (A5b) $\beta$ is real and positive, one has

$$\text{Prob}[\,|1_c, 0, 0\rangle\,] = [|\alpha(j_0, k_0)|^2 + \beta]^2 = \beta^2 + 2|\alpha(j_0, k_0)|^2 \beta + |\alpha(j_0, k_0)|^4.$$

The interference term $2|\alpha(j_0, k_0)|^2 \beta$ is positive, s.t. on the path $c$ appears constructive interference between the UV-beam incident on the crystal and the up-conversion photons resulting from the incident pairs, see (A10). The result is an increased up-conversion rate. Replacing $\beta^2$ according to (A8) there results

$$\text{Prob}[\,|1_c, 0, 0\rangle\,] = = 1 + |\alpha(j_0, k_0)|^2 - \sum_{\substack{j \neq j_0 \\ k \neq k_0}} |\alpha(j, k)|^2 - 2|\alpha(j_0, k_0)|^2 (1 - \beta) + |\alpha(j_0, k_0)|^4.$$

A quite lengthy but simple calculus shows that the last three terms account for the tiny amount of selected pairs not up-converted in $X$, and for the down-conversion of UV-photons to non-selected pairs

(A12) $\text{Prob}[\,|1_c, 0, 0\rangle\,] = 1 + |\alpha(j_0, k_0)|^2 - |\alpha(j_0, k_0)\, \gamma(j_0, k_0)|^2 - \sum_{\substack{j \neq j_0 \\ k \neq k_0}} |\alpha(j, k) + \alpha(j_0, k_0)\, \gamma(j, k)|^2.$

The third term represents the probability that a selected pair pass through the crystal without being up-converted. Looking at the expression of $\gamma(j_0, k_0)$ in (A11), it is obvious that this term is even smaller than $|\alpha(j_0, k_0)|^6$, s.t. may be neglected. Multiplying (A12) by $N_0$ one finds that the rate of UV-photons on $c$, $\langle Q_c \rangle$, satisfies the relation

(A13) $N_0 - \langle Q_c \rangle \approx N_0 \sum_{\substack{j \neq j_0 \\ k \neq k_0}} \text{Prob}[\,|0, 1_j, 1_k\rangle\,] - N_0 |\alpha(j_0, k_0)|^2.$

$N_0$ is the rate of the UV-photons incident on the crystal $X$ through the path $p$, so, the LHS represents the difference between the UV-rate exiting the crystal and the UV-rate incident to the crystal. As the RHS shows, this rate increases in $X$ by a quantity that comparing with (6) is just equal to $\langle Q' \rangle$, the rate of selected pairs coming from $X'$. The second term on the RHS is the loss of UV-photons on creation of non-selected pairs. The main conclusion of this case comes from the first term on RHS: for $\sigma = \varphi - \pi/2$, the crystal $X$ up-converts all the selected pairs coming from $X'$.

b) For $\sigma = \varphi$, the production of selected pairs in $X$ is enhanced as shown in the text, eq. (11). From this, a part $\langle Q' \rangle = \frac{1}{4} \langle Q_E \rangle$ is contributed by what comes from the crystal $X'$. The rest of $\frac{3}{4} \langle Q_E \rangle$ originates in the crystal $X$. That can be shown directly by calculating the rate of photons on the path $c$ and comparing it with the rate on $p$.

Introducing in (10) the relation $\sigma = \varphi$, one gets the probability of having a UV-photon on the path $c$

$$\text{Prob}[|1_c,0,0\rangle] = [-|\alpha(\boldsymbol{j}_0,\boldsymbol{k}_0)|^2 + \beta]^2 = \beta^2 - 2|\alpha(\boldsymbol{j}_0,\boldsymbol{k}_0)|^2\beta + |\alpha(\boldsymbol{j}_0,\boldsymbol{k}_0)|^4.$$

Since according to (A5b) $\beta$ is positive, the interference term $-2|\alpha(\boldsymbol{j}_0,\boldsymbol{k}_0)|^2\beta$ is negative, so that destructive interference appears between the UV-photons incident on the crystal and the up-conversion photons, see (A10). Replacing $\beta^2$ according to (A6), the above equality becomes

$$(A14)\quad \text{Prob}[|1_c,0,0\rangle] = 1 - 3|\alpha(\boldsymbol{j}_0,\boldsymbol{k}_0)|^2 - \sum_{\substack{\boldsymbol{j}\neq\boldsymbol{j}_0 \\ \boldsymbol{k}\neq\boldsymbol{k}_0}}|\alpha(\boldsymbol{j},\boldsymbol{k})|^2 + 2|\alpha(\boldsymbol{j}_0,\boldsymbol{k}_0)|^2(1-\beta) + |\alpha(\boldsymbol{j}_0,\boldsymbol{k}_0)|^4.$$

The last two terms are very small comparing to the others, and may be neglected. Multiplying by $N_0$, one finds that the photon rate on $c$, $\langle Q_c \rangle$, satisfies the relation

$$(A15)\quad N_0 - \langle Q_c\rangle \approx 3N_0|\alpha(\boldsymbol{j}_0,\boldsymbol{k}_0)|^2 + N_0\sum_{\substack{\boldsymbol{j}\neq\boldsymbol{j}_0 \\ \boldsymbol{k}\neq\boldsymbol{k}_0}}|\alpha(\boldsymbol{j},\boldsymbol{k})|^2.$$

$N_0$ is the rate of the UV-photons on the path $p$. As the RHS of (A15) shows, inside the crystal $X$ this rate decreases due to down-conversion to selected and non-selected pairs. Comparing the first term on the RHS with (6) one finds that for $\sigma = \varphi$ the rate of down-conversion to selected pairs in $X$ is tripled in comparison with the rate in the same crystal in the experiment **A**.

c) In general, from eqs. (7) and (10) one gets

$$N_0\langle\Phi|\Phi\rangle = \langle Q'\rangle + N_0 = \langle Q_c\rangle + N_0\text{Prob}[|0,1_{\boldsymbol{j}_0},1_{\boldsymbol{k}_0}\rangle] + N_0\sum_{\substack{\boldsymbol{j}\neq\boldsymbol{j}_0 \\ \boldsymbol{k}\neq\boldsymbol{k}_0}}\text{Prob}[|0,1_{\boldsymbol{j}},1_{\boldsymbol{k}}\rangle].$$

therefore, by rearranging terms,

$$(A16)\quad N_0 - \langle Q_c\rangle = N_0\text{Prob}[|0,1_{\boldsymbol{j}_0},1_{\boldsymbol{k}_0}\rangle] + N_0\sum_{\substack{\boldsymbol{j}\neq\boldsymbol{j}_0 \\ \boldsymbol{k}\neq\boldsymbol{k}_0}}\text{Prob}[|0,1_{\boldsymbol{j}},1_{\boldsymbol{k}}\rangle] - \langle Q'\rangle.$$